\documentclass[%
preprintnumbers,
nofootinbib,
 amsmath,amssymb,
 aps,
]{revtex4}

\usepackage{epsfig}
\usepackage{graphicx}
\usepackage{mathrsfs}
\usepackage{amsfonts}
\usepackage{caption2}
\usepackage{array}
\begin{document}

\title{Torsion Cosmology of Poincar\'e gauge theory and the constraints of its
  parameters via SNeIa data}
\author{Xi-Chen Ao}
\author{Xin-Zhou Li}

\affiliation{Shanghai United Center for Astrophysics(SUCA),
 Shanghai Normal University, 100 Guilin Road, Shanghai 200234,China}

\email{aoxichen@gmail.com, kychz@shnu.edu.cn}

\begin{abstract}
Poincar\`e gauge theory (PGT) is an alternative gravity theory, which attempts to
bring the gravity into the gauge-theoretic frame, where the Lagrangian is quadratic
in torsion and curvature. Recently, the cosmological
models with torsion based on this theory have
drawn many attentions, which try to explain the cosmic acceleration in a
new way. Among these PGT cosmological models, the one with only even parity
dynamical modes -- SNY model, for its realistic meaning, is very attractive.
In this paper, we first analyze the past-time cosmic evolution of SNY
model analytically. And based on these results we fit this model to
the most comprehensive SNeIa data (Union 2) and thus find the best-fit values of model
parameters and initial conditions, whose related $\chi^{2}$ value is consistent
with the one from $\Lambda$CMD at the 1$\sigma$ level. Also by the $\chi^{2}$
estimate, we provide certain constraints on these parameters. Using these best-fit
values for the Union 2 SNeIa dataset, we are able to predict the evolution
of our real universe over the late time. From this prediction, we know the fate of our universe that
it would expand forever, slowly asymptotically to a halt, which is in accordance
with the earlier works.
\end{abstract}


\maketitle

\section{\label{sec:introduction}Introduction}
About 12 years ago, the high redshift SNeIa observations, suggesting that our
universe is not only expanding but also accelerating, have changed  our  basic  understanding
of the universe, which led us to the new era of cosmology.
In order to explain this weird phenomena, cosmologists introduced the cosmological
constant back again, and constructed a new standard model also known as the
concordance model,  which is intended to satisfy
 all the main observations such as type Ia supernovae (SNeIa), cosmic microwave
 background radiation (CMBR) and large scale
structure (LSS). Although the cosmological constant accounts for almost 74\% of the
whole energy density in this  concordance model, the value is still too small to
be explained by any current fundamental theories. Lacking the underlying
theoretical foundations, the particular value of cosmological constant
is just selected phenomenologically, which means the model is highly  sensitive
to the value of model parameter, resulting in the so-called the fine-tuning
problem. This problem is considered as the biggest issue for almost all cosmological models.
In order to alleviate this troublesome problem, various dynamical dark energy theories have
been proposed and developed these years, such as quintessence \cite{Peebles:2002gy,Li:2001xaa}
and phantom \cite{Caldwell:1999ew,Li:2003ft},
in which the energy composition depends on time. But these exotic fields are still
phenomenological, lacking theoretical foundations. Besides adding
some unknown fields, there is another kind of theories  known as
modified gravity,  which use alternative gravity theory instead of Einstein theory,
such as $ f(R)$ theory \cite{Nojiri:2010wj,Du:2010rv}, MOND
cosmology \cite{Zhang:2011uf}, Poincar\'e gauge theory \cite{Hehl:1994ue,Blagojevic2002}, and de Sitter gauge
theory \cite{Ao:2011dn}. Among these
theories,  Poincar\'e gauge theory (PGT) has a solid theoretical
motivation, which tries to bring the gravity into the gauge-theoretic framework.

The attempt to treat gravitation as a gauge interaction could date back to
the 1950s. Utiyama first presented a groundwork for a gauge theory of
gravitation \cite{Utiyama:1956sy}. Later
Kibble and Sciama \cite{Kibble:1961ba} inherited Utiyama's work and developed a gauge
theory of gravitation commonly called Einstein-Cartan-Sciama-Kibble (ECSK)
theory, where the symmetry group is Poincar\'e group.
In ECSK theory, the Lagrangian has only one linear curvature term, thus the
equation of torsion is algebraic, and torsion is
non-dynamical.  Some follow-up cosmological models with torsion, based on ECSK model, were investigated,
which try to avoid the cosmological singularity \cite{Kerlick:1976uz}, and from
these cosmological discussions, they found the torsion
was imagined as playing role only at high densities in the early universe.
In order to enable  torsion to propagate, one has to introduce quadratic term of
torsion and curvature, which is spoken of Poincar\'e gauge theory, proposed by
Hehl in 1980 \cite{Hehl1980}.  Poincar\'e group is a semiproduct of translation
group and Lorentz rotation group, which is the global symmetry group of a space-time
in absence of gravity. If one localize this global symmetry as a gauge
symmetry, the gravity would emerge spontaneously. Note that in this
transition some independent compensating fields have to be introduced, the orthonormal coframe and
the metric-compatible connection, which correspond to the translation and  local
rotation potentials, respectively. And their corresponding field strength are torsion and
curvature. The early works of PGT are
intended to unify gravity and the other three interactions in the framework of
gauge field theory. However now it might be the solution to the cosmic
acceleration problem, in which the dynamical torsion could play the role as dark
energy to drive our universe to accelerate\cite{Shie:2008ms,Chen:2009at,Ho:2011qn,Ho:2011xf}.

In PGT, the propagating torsion field can be
identified as six possible dynamic modes, carrying spins and
parity: $2^{\pm}, 1^{\pm},0^{\pm}$, respectively \cite{Hayashi:1979wj,Hayashi:1980ir}. However, expect for two
"scalar modes", $0^{\pm}$, other dynamical modes are not physically acceptable,
which are ruled out by certain theoretical constraints (e.g., "no-ghosts" and
"no-tachyons"). The pseudoscalar mode with the odd parity, $0^{-}$, is reflected in the
axial vector torsion, which is driven by the intrinsic spin of
elementary fermions. 
It is generally thought that the axial torsion is large and has notable
effects only in the very early universe, where the spin density is large.
Therefore, this part probably have insignificant contribution to the current
evolution of our universe. Whereas the scalar mode with the even parity,
$0^{+}$, which is associated with the vector torsion, does not interact in any direct way with any known
of matter. Consequently, it is rational to imagine it has an  significant
magnitude but yet not been noticed.  In 2008, Shie, Nester and Yo (SNY) proposed a
cosmological model based on PGT, where the dynamical scalar torsion $0^{+}$ accounts for the cosmic
acceleration \cite{Shie:2008ms,Chen:2009at}. Later, we investigated the dynamical properties of
SNY model and applied the statefinder diagnostics to
it\cite{Li:2009zzc,Li:2009gj}, and did some analytical analysis on the late-time
evolution in \cite{Ao:2010mg}.  In 2010,
Baekler, Hehl and Nester generalized the SNY model
to the BHN model which includes the $0^{-}$ mode and even the cross parity
couplings and gave a comprehensive picture of PGT
cosmology \cite{Baekler:2010fr}. And some follow-up works suggest the cosmic acceleration
is still mainly due to the even mode. \cite{Ho:2011qn,Ho:2011xf}

In this paper, we study the evolution of our universe based on SNY model.
In SNY model, the cosmic evolution is described in terms of three variables,
($H, \Phi, R$), by three first-order dynamical equations. We obtain three
exact solutions of scale factor with certain particular constant affine scalar curvature, and show that
these solutions are not physically real, which requires us to study the non-constant curvature case.
Then we extend the investigation to the non-constant affine scalar curvature, and find the analytical solutions
for the past-time evolution, which could be compared with the observational data
to constrain the values of parameters and determine the goodness of this model.
Then we fit these theoretical results to the Union 2 supernovae dataset, and find the best-fit values of the model parameters
and initial conditions, by means of the $\chi^{2}$ estimate. Also, we plot the
contours of certain confidence levels, which shows some constraints on the
parameters. Finally, we study the future evolution. Using these constraints of parameters, we predict that the real universe
would expand forever, slowly asymptotically to a halt, which is consistent with
some earlier works.\cite{Li:2009gj, Ao:2010mg}

\section{\label{sec:poinc-gauge-theory}SNY model and its some exact solutions}
In order to localize the global Poincar\'e symmetry, one has to introduce two
compensating fields: coframe field $\vartheta^{\alpha}$ and metric-compatible
connection field
$\Gamma^{\alpha\beta}=\Gamma^{[\alpha\beta]}_{i}\mathrm{d}x^{i}$ \footnote{the Greek indices $\alpha,\beta, \gamma...$ denote the 4d orthonormal indices, whereas the Latin indices denote 4d coordinate (holonomic) indices and $i,j,k,...$ denotes }. And their
associated field strengths are the torsion and curvature 2-forms
\begin{eqnarray}
  \label{eq:gauge-field-strength1}
  T^{\mu} &\equiv& \frac{1}{2}T^{\mu}_{\alpha\beta}~ \vartheta^{\alpha}\wedge
  \vartheta ^{\beta}=\mathrm{d}\vartheta ^{\mu} +\Gamma^{\mu}_{\nu}\wedge
  \vartheta ^{\nu}\\
\label{eq:gauge-field-strength2}
  R^{\mu\nu}&\equiv&
  \frac{1}{2}R^{\mu\nu}_{\alpha\beta}~\vartheta^{\alpha}\wedge\vartheta^{\beta}=\mathrm{d}
  \Gamma^{\mu\nu}+\Gamma^{\mu}_{\rho}\wedge\Gamma^{\rho\nu}
\end{eqnarray}
which satisfy the respective Bianchi identities:
\begin{eqnarray}
  \label{eq:bianchi}
  \mathrm{D}T^{\mu}\equiv R^{\mu}_{\, \nu}\wedge \vartheta^{\nu},\quad
\mathrm{D} R^{\mu}_{\, \nu}\equiv0
\end{eqnarray}
The Lagrangian denstiy in PGT takes the standard quadratic Yang-Mills form, qualitively,
\begin{eqnarray}
  \label{eq:lagrangian-1}
  \mathcal{L}[\vartheta,
  \Gamma]\sim \mathrm{curvature} +(\mathrm{ torsion})^{2}+(\mathrm {curvature})^{2}.
\end{eqnarray}

Some early works on PGT concluded that in the weak field approximation, the
torsion field could be identified as six irreducible parts with certain
dynamical modes, which propagate with  $2^{\pm}$, $1^{\pm}$,
$0^{\pm}$. Later some investigations by linearized theory and Hamiltonian
analysis concluded that only the two "scalar modes" are physically acceptable,
carrying $0^{\pm}$. Furthermore, because the $0^{-}$ is driven by spin
density, this mode would not have a significant effects, expect in the very
early universe. And some numerical results demonstrate that the acceleration
is mainly due to the $0^{+}$. So here we only consider the simple
 $0^{+}$ case, i.e., SNY model. In this case, the gravitational lagrangian density is
simplified to this specific form,
\begin{eqnarray}
  \label{eq:lagrangian-2}
  \mathcal{L}[\vartheta,
\Gamma]=\frac{1}{2\kappa}\left[-a_{0}R+\sum^{3}_{n=1}a_{n}{\buildrel
(n)\over T}{}^{2}+\frac{b}{12}R^{2}\right],
\end{eqnarray}
where is the algebraically irreducible parts of torsion, R is the scalar curvature, which is the non-vanishing
irreducible parts of curvature \cite{Baekler:2010fr}. Note that $a_{0}$ and
$a_{n}$ are dimensionless parameters, whereas $b$ have the same dimension
with $R^{-1}$.

Since current observations favor a homogeneous, isotropic and spatially flat universe, it is reasonable to
work on the Friedman-Lemaitre-Robertsen-Walker (FLRW) metric, where the  isotropic orthonormal
coframe takes the form:
\begin{eqnarray}
  \label{eq:coframe}
  \vartheta ^{0}= \mathrm{d}t, \qquad \vartheta^{i}=a(t) \mathrm{d}x^{i};
\end{eqnarray}
and the only non-vanishing connection 1-form coefficients are of the form:
\begin{eqnarray}
 \label{eq:connection}
  \Gamma^{i}_{0}=\Psi(t)\mathrm{x}^{i},
\end{eqnarray}
Consequently, the nonvanishing torsion tensor components take the form
\begin{eqnarray}
  \label{eq:torsion-tensor}
  T^{i}_{~j0}=a(t)^{-1}(\Psi(t)-\dot{a}(t))\delta^{i}_{j}\equiv \frac{-\Phi(t)}{3}\delta^{i}_{j}
\end{eqnarray}
\textit where $\Phi$ represents torsion.
By variation w.r.t. the coframe and connection, one could find the cosmological equations of SNY model.
\begin{eqnarray}
  \label{eq:motion-equation-H}
  \dot H &=& \frac{\mu}{6 a_{2}}R-\frac{\kappa\;\rho_{m}}{6 a_{2}}-2
H^{2}\\
  \label{eq:motion-equation-phi}
\dot \Phi &=& \frac{a_{0}}{2a_{2}}R-\frac{\kappa\;\rho_{m}}{2
a_{2}}-3H\Phi +\frac{1}{3}\Phi^{2}\\
  \label{eq:motion-equation-R}
\dot R &=& -\frac{2}{3}\left(R+\frac{6\mu}{b}\right)\Phi,
\end{eqnarray}
where the $H=\dot{a}/a$ is the Hubble parameter, $\mu=a_{2}-a_{0}$,  which represents the
mass of  $0^{+}$ mode, and the energy density of matter component is
\begin{eqnarray}
  \label{eq:energy-density}
\kappa
\rho_{m}=\frac{b}{18}\left(R+\frac{6\mu}{b}\right)(3H-\Phi)^{2}-\frac{b}{24}R^{2}-3a_{2}H^{2}.
\end{eqnarray}
Note that the universe here is assumed to
be the dust universe, for the effect of radiation is almost negligible in the
late-time evolution. The Newtonian limit requires $a_{0}=-1$.

From Eq.\eqref{eq:motion-equation-R} , it is easy to find the scalar affine curvature remains
a constant $R = -6\mu/b$ forever as long as its initial data has
this special value. In this case, Eq.~\eqref{eq:motion-equation-H} can be
rewritten as
\begin{equation}
2a\ddot{a}+\dot{a}^{2}+\frac{3}{2}\frac{\mu^{2}}{a_{2}b}a^{2}=0,
\end{equation}
where the scale factor $a(t)$ is decoupled to $\Phi$ and $R$ fields. Thus, we could
obtain some simple exact solution of $a(t)$.

The positivity of the kinetic energy requires $a_2 > 0$ and $b > 0$ \cite{Yo:1999ex}, so we have the solution
\begin{equation}
\label{eq:sol-1}
a(t)=a_{0}\left(\frac{\cos\left[\frac{3\zeta}{2}\left(t-t_{0}\right)-\arctan\left(\frac{H_{0}}{\zeta}\right)\right]}
{\cos\left[\arctan\left(\frac{H_{0}}{\zeta}\right)\right]} \right)^{\frac{2}{3}},
\end{equation}
where $\zeta = \mu/\sqrt{2a_2b}$ and $H_0 = H(t_0)$. However, such a
choice conflicts with the assumption of energy positivity in the $R
= -6\mu/b$ case.

If we audaciously relax the parameter requirement for positive
kinetic energy, \textit{i.e.}, $a_2 < -1$ and $\mu < 0$, this
phantom scenario will turn out to be interesting. Now we have the
solution
\begin{eqnarray}
  \label{eq:sol-2}
 & a(t)&=a_{0}\exp\left[(\xi-2)(t-t_{0})+\frac{2}{3\xi}\ln \frac{(\xi-H_{0})+(\xi+H_{0})\exp[3\xi(t-t_{0})]}{2\xi}\right],
\end{eqnarray}
where $\xi = \mu/\sqrt{-2a_2b}$.
From this expression, it is obvious that the late-time behavior would be analogous with the exponential
expansion of  the de Sitter universe. In other words, dark energy can be
mimicked in such case.
Using the dynamical analysis, we have pointed out that there is a
late-time de Sitter attractor \cite{Li:2009zzc,Ao:2010mg}. Note that the solution \eqref{eq:sol-2}
is just corresponding to the de Sitter attractor.

Especially, as $a_2 = -1$, we have a solution
\begin{equation}
a(t) = a_0\left(\frac{2+3H_0t}{2+3H_0t_0}\right)^{\frac{2}{3}}.
\end{equation}
From this expression, we could find that the late-time evolution is
similar to the $t^{2/3}$ expansion of  the matter dominant universe.

In this constant affine scalar
curvature case, the three exact solutions we find are all not physically
acceptable: the first conflicts with the assumption of energy positivity; while
the second and third ones breaks the positivity of kinetic energy. For this
reason, these solutions with constant scalar curvature cannot describe our real
universe, and thus we have to study the case of non-constant scalar curvature.

Unfortunately, for  such a complex nonlinear system,
Eqs.~\eqref{eq:motion-equation-H}-\eqref{eq:motion-equation-R},
the whole evolution of our universe is difficult to obtain, so we are forced to
divide it into two seperate parts, the past and the future, and discuss them
respectively. Among these two parts, it is the past one that could be compared with the existing
observational results and impose constraints on the values of model parameters and
initial conditions.
Based on these constraints, we could obtain the future
evolution and the fate of our real universe.
Therefore, we study the past-time evolution first.

\section{The Analytical Approach for the $a(t)<a(t_{0})$}
\label{sec:Past}
In Ref.\cite{Ao:2010mg}, we analyzed the evolution when $a(t)>a(t_{0})$,
i.e. the future evolution. In this section, we investigate the evolution when $a(t)<a(t_0)$, i.e. the evolution of
universe at $t<t_{0}$.

For such a dust universe, the continuity equation is
\begin{eqnarray}
  \label{eq:continuity-equation}
  \dot{\rho}=-3H\rho,
\end{eqnarray}
which could also be derived directly by the dynamical equations
Eqs.\eqref{eq:motion-equation-H}-\eqref{eq:motion-equation-R} and the constraint
equation Eq.~\eqref{eq:energy-density}.
This equation is easy to solve,
\begin{eqnarray}
  \label{eq:rho}
  \rho=\frac{\rho_{0}}{a^{3}},
\end{eqnarray}
where $\rho_{0}$ is the current matter density of our universe.

If we  rescale the variables and parameters as
\begin{eqnarray}
&&t\rightarrow t/l_0;\quad H\rightarrow l_0 H;\quad  R\rightarrow l_0^{2} R;\nonumber \\
&&\Phi\rightarrow l_0 \Phi;\quad \rho \rightarrow l_{0}^{2}\kappa \rho;\quad
b\rightarrow b/l_{0}^{2},\label{transformation}
\end{eqnarray}
where $l_0=c/H_0$ is the Hubble radius,  these motion equations
Eqs.\eqref{eq:motion-equation-H}-\eqref{eq:motion-equation-R} would be
dimensionless.

By the substitution Eq.\eqref{eq:rho} back to
Eqs.\eqref{eq:motion-equation-H}-\eqref{eq:motion-equation-R} and rescaling
transformations, one could obtain the new equation set, which is more convenient
to solve,
\begin{eqnarray}
  \label{eq:modified-motion-equation-h}
  a H H'&=&\frac{\mu}{6 a_{2}}R- \frac{\rho_{0}}{6
    a_{2}}\frac{1}{a^{3}}-2H^{2}\\
a H \Phi'&=&\frac{1}{2 a_{2}}R-\frac{\rho_{0}}{2 a_{2}} \frac{1}{a^{3}}
-3H \Phi + \frac{1}{3} \Phi^{2}\\
  \label{eq:modified-motion-equation-r}
aH R'&=&-\frac{2}{3}(R+\frac{6\mu}{b})\Phi,
\end{eqnarray}
where the prime means the derivation with respect to scale factor $a$.
The current scale factor is generally supposed to be unity,
so it is reasonable to assume the ans$\ddot{\mathrm{a}}$tz is as follows,
\begin{eqnarray}
  \label{eq:ansatz-h}
 H(a)&=&h_{0}+ \sum^{\infty}_{n=1}h_{n}(a-1)^{n},\\
\Phi(a)&=& \varphi_{0}+\sum^{\infty}_{n=1}\varphi_{n}(a-1)^{n},\\
\label{eq:ansatz-r}
R(a)&=&r_{0}+ \sum^{\infty}_{n=1}r_{n}(a-1)^{n},
\end{eqnarray}
which has a good convergence when $a(t)<a(t_{0})$.\footnote{Here, $a(t)$ is set
  to unity.}

Substitute this ans$\ddot{\mathrm{a}}$tz back to the
Eqs.\eqref{eq:modified-motion-equation-h}-\eqref{eq:modified-motion-equation-r},
we find the recursion relation of coefficients,
\begin{eqnarray}
  \label{eq:n=0-h}
 && h_{1}=\frac{1}{h_{0}}\left(\frac{\mu r_{0}}{6 a_{2}}-\frac{ \rho_{0}}{6 a_{2}}-2h_{0}^{2}\right),\\
&&\varphi_{1}=\frac{1}{h_{0}}\left(\frac{r_{0}}{2a_{2}}-\frac{\rho_{0}}{2a_{2}}-3h_{0}\varphi_{0}
+\frac{\varphi_{0}^{2}}{3}\right),\\
&&r_{1}=-\frac{2}{3h_{0}}\left(r_{0}+\frac{6\mu}{b}\right)\varphi_{0},
\end{eqnarray}
when $n\ge 2$,
\begin{eqnarray}
  \label{eq:n>0-h}
  h_{n}&=&\frac{1}{n h_{0} }\left[\frac{\mu r_{n-1}}{6a_{2}}-\frac{\rho_{0}}{12a_{2}}n(n+1)-2\sum^{n-1}_{i=0}h_{i}h_{n-i}-\sum^{n-1}_{i=0}(n-1-i)h_{i}h_{n-1-i}
\right.\nonumber\\
&&\left.-\sum^{n-1}_{i=1}(n-i)h_{i}h_{n-i}\right],\\
  \label{eq:n>0-phi}
\varphi_{n}&=&\frac{1}{nh_{0}}\left[\frac{r_{n-1}}{2a_{2}}-\frac{\rho_{0}}{4a_{2}}n(n+1)
-3\sum^{n-1}_{i=0}h_{i}\varphi_{n-1-i}+\sum^{n-1}_{i=0}\frac{\varphi_{i}\varphi_{n-1-i}}{3}\right.\nonumber\\
&&\left.
-\sum^{n-1}_{i=0}(n-1-i)h_{i}\varphi_{n-i}
-\sum^{n-1}_{i=1}(n-i)h_{i}\varphi_{n-i}\right],\\
  \label{eq:n>0-r}
r_{n}&=&\frac{1}{n h_{0}}\left[\frac{2}{3}\sum^{n-1}_{i=0}r_{i}\varphi_{n-1-i}+\frac{4\mu}{b}\varphi_{n}\sum^{n-1}_{i=1}(n-i)h_{i}r_{n-i}\right],
\end{eqnarray}
with
\begin{eqnarray}
  \label{eq:initial-rho}
  \rho_{0}=\frac{b}{18}\left(r_{0}+\frac{6\mu}{b}\right)(3h_{0}-\varphi_{0})^{2}-\frac{b}{24}r_{0}^{2}-3a_{2}h_{0}^{2},
\end{eqnarray}
where $h_{0},~\phi_{0}$ and $ r_{0}$ are the initial conditions of $H(1),~\Phi(1)$ and $R(1)$,
i.e. the present time. Note that for the rescaling transformation Eq.\eqref{transformation}, the initial
value for $H$ is supposed to be unity, and therefore $h_{0}=1$. We here obtained the
analytical solution of the past evolution  of
SNY model, which can be  tested by the observational results, realistically.
\section{The Constraints of Parameters via SNeIa}
\label{sec:fitting}
The most common approach to test a cosmological model is the supernovae
fitting. In this section we attempt to fit the model parameters and the initial
values to the current type Ia supernovae data.

The supernovae data we use here is the Union 2 dataset (N=557), which is the most
comprehensive one up to date, combining the former SNeIa dataset in a
homogeneous manner. It consists of  distance modulus $\mu_{obs}$, which equals
to the difference between apparent magnitude $m_{i}$ and the absolute magnitude
$M_{i}$, redshifts $z_{i}$ of supernovae, and the covariance matrix $C_{SN}$ represents
the statistical and systematic errors. By comparing the theoretical distance modulus $\mu_{th}$, derived from the
cosmological model,  with the observational data, we could obtain the
constraints on the model parameters and initial value as well as the goodness of
the model.

As stated above, the SNY model predicts a specific form of the Hubble
constant $H(a; a_{2},b,\varphi_{0},r_{0})$ as a
function of scale factor, which is explicitly expressed in
Eqs.\eqref{eq:ansatz-h},\eqref{eq:n=0-h} and \eqref{eq:n>0-h}. For the
relation between scale factor $a$ and redshift $z$, it is easy to rewrite
$H(a)$ in terms of redshift,
\begin{eqnarray}
  \label{eq:hubble-constant}
  H(z; a_{2},b,\varphi_{0},r_{0})=\sum^{\infty}_{n=0}(-1)^{n}h_{n}\left(\frac{z}{1+z}\right)^{n},
\end{eqnarray}
which is more convenient to compare to the supernovae data.

The theoretical distance modulus is related to the luminosity distance $d_{L}$ by
\begin{eqnarray}
  \label{eq:modulus}
  \mu_{th}(z_{i})&=&5 \log_{10}\left(\frac{d_{L}(z_{i})}{\mathrm{Mpc}}\right)+25 \nonumber\\
&=&5 \log_{10}D_{L}(z_{i})-5\log_{10}\left( \frac{c H_{0}^{-1}}{\mathrm{Mpc}} \right)+25 \nonumber\\[0.12cm]
&=& 5 \log_{10}D_{L}(z_{i})-5\log_{10}h +42.38,
\end{eqnarray}
where the $D_{L}(z)$ is the dimensionless 'Hubble-constant free' luminosity
distance defined by $D_{L}(z)=H_{0}d_{L}(z)/c$.

\begin{figure}
\centering
\includegraphics[width=8.5cm]{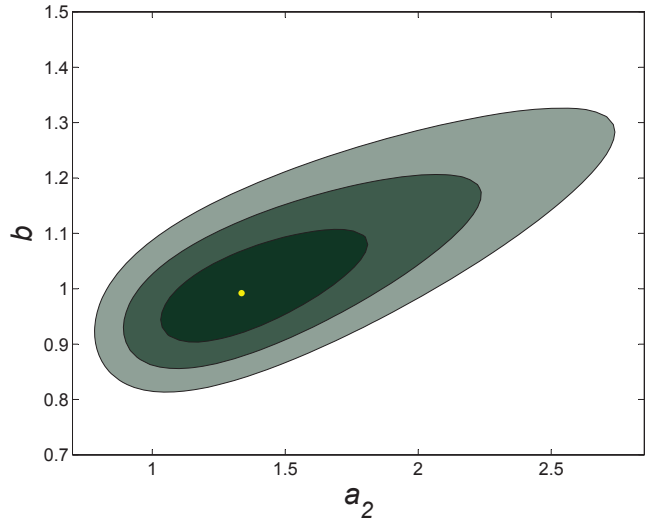}
\caption{\label{fig:mpcz} The 68.3\%, 95.4\% and 99.7\% confidence
  contours with respect to model parameters $a_{2},~b$, using the Union 2
  dataset. Here we assume the initial conditions are $\varphi_{0}=0.584$ and
  $r_{0}=5.839$. The yellow point denotes the best-fit point.}
\end{figure}

\begin{figure}
\centering
\includegraphics[width=8.5cm]{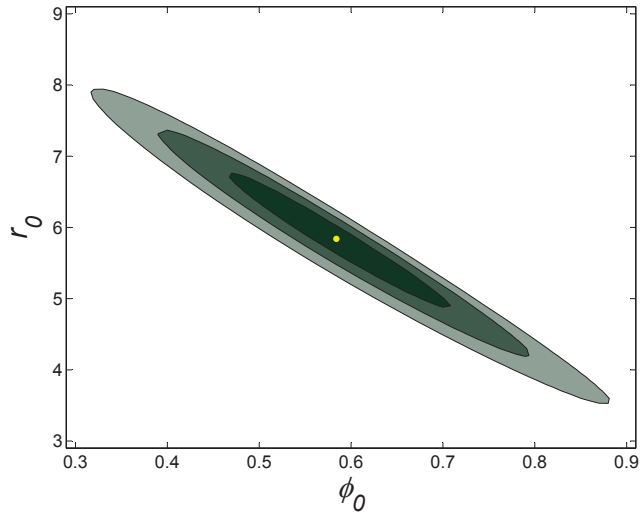}
\caption{\label{fig:ivcz}The 68.3\%, 95.4\% and 99.7\%  confidence
  contours with respect to the initial values of $\varphi_{0},~r_{0}$, using the Union 2
  dataset. Here we assume the model parameters are $\varphi_{0}=0.584$ and
  $r_{0}=5.839$. The yellow point denotes the best-fit point.}
\end{figure}

For the spatially flat model we consider here, the 'Hubble-constant free'
luminosity distance could be expressed in terms of Hubble parameter
$H(z;a_{2},b,\phi_{0},r_{0})$, Eq.\eqref{eq:hubble-constant} ,
\begin{eqnarray}
  \label{eq:dL}
D_{L}(z)&=& (1+z)\int^{z}_{0} \mathrm{d}z' \frac{1}{H(z'; a_{2},b,\phi_{0},r_{0})}.
\end{eqnarray}
Due to the normal distribution of errors, the $\chi^{2}$ method could be used as
the maximum likelihood estimator to compare the theoretical models
and the observational data, which would show us the best-fit parameters ($a_{2},
b, \varphi_{0}$ $,r_{0}$) and the goodness of model. The $\chi^{2}$ here for the
SNeIa data is
 \begin{eqnarray}
  \label{eq:chi22}
 \chi^{2}(\theta)&=&\sum^{N}_{i,j}\left[\mu_{obs}(z_{i})-\mu_{th}(z_{i})\right]
 (C_{SN}^{-1})_{i j}[\mu_{obs}(z_{j})-\mu_{th}(z_{j})]\nonumber \\
&=&\sum^{N}_{i,j}\left[\mu_{obs}(z_{i})-5\log_{10}D_{L}(z_{i};\theta)-\mu_{0}\right] (C_{SN}^{-1})_{i j}[\mu_{obs}(z_{j})-5\log_{10}D_{L}(z_{i};\theta)-\mu_{0}].
\end{eqnarray}
where $\mu_{0}=-5\log_{10}h + 42.38$ and $\theta$ denotes the model parameters and
initial values. The parameter $\mu_{0}$ is a nuisance parameter, whose
contribution we are not interested in. By marginalization over this parameter
$\mu_{0}$, we obtain
\begin{eqnarray}
 \label{eq:chi23}
\tilde{\chi}^{2}(\theta)=A(\theta)-\frac{B(\theta)^{2}}{C}+\ln\left(\frac{C}{2\pi}\right),
\end{eqnarray}
where
\begin{eqnarray}
 \label{eq:marginalization}
A(\theta)&=&\sum^{N}_{i,j}\left[\mu_{obs}(z_{i};\theta)-5\log_{10}D_{L}(z_{i};\theta)\right](C_{SN}^{-1})_{i
   j}[\mu_{obs}(z_{j};\theta)-5\log_{10}D_{L}(z_{i};\theta)],\\
B(\theta)&=&\sum^{N}_{j} (C_{SN}^{-1})_{i   j}[\mu_{obs}(z_{j};\theta)-5\log_{10}D_{L}(z_{i};\theta)],\\
C(\theta)&=&\sum^{N}_{i,j}=(C_{SN}^{-1})_{i,j}.
\end{eqnarray}
By minimizing the $\chi^{2}$,
we find the best-fit values of parameters and
initial conditions ($a_{2}, b, \varphi_{0}, r_{0}$) of SNY model \footnote{ The full numerical calculation
are performed using Matlab here.}, as shown in Tab.\ref{tab:best-fit}.
\begin{table}[h]
\centering
\begin{tabular}{|p{0.3cm}p{1cm} p{1cm} p{1.cm}p{1.cm}|p{1.5cm}|}\hline
& $a_{2}$  &$b$ &$\varphi_{0}$&$r_{0}$&~$\chi^{2}$\\
\hline
&&&&&\\[-0.25cm]
& 1.336 & 0.992  &0.584   &5.839   &~535.284\\
\hline
\end{tabular}
\caption{\label{tab:best-fit}The best-fit  initial data and parameters }
\end{table}
The minimal $\chi^{2}$ here is $535.284$ , whereas the value for $\Lambda$CDM is
$536.634$, with $\Omega_{m}=0.27$, $\Omega_{\Lambda}=0.73$, which implies that
the torsion cosmology is consistent with $\Lambda$CMD at the 1$\sigma$ level. Substitute these
best-fit values back to \eqref{eq:initial-rho}, we obtain the initial value of
$\rho_{0}=1.004$.
Since the parameter space here is 4-dimensional, we cannot
show the contour of confidence level in one picture. We have to analyze it in two
"cases". First, we fix the initial values at their best-fit values, and obtain
the corresponding contours of some particular confidence levels of the
model parameters, as shown in Fig.\ref{fig:mpcz}. Second, we fix the model parameter at their best-fit
values, and obtain the contours of confidence levels of initial
conditions, as shown in Fig.\ref{fig:ivcz}. From these two contours, it is easy
to find that, to some extent, the sensitivity of model parameters and initial
conditions has been lowered in torsion cosmology.

\section{Fate of the Universe}
\label{sec:future}

\label{sec:analytical-late-time}

The fate of the universe is an essential issue, which is discussed widely, for almost every cosmological model.
In PGT cosmology, many works have also been
conducted. In \cite{Shie:2008ms,Chen:2009at}, some numerical analyses have been
done, which showed that $H$, $\Phi$ and $R$ have a periodic
character at late-time of the evolution for $a_{2}>0$ and $b>0$,
approximately. Some follow-up  dynamics analysis and statefinder diagnostic done  in \cite{Li:2009zzc, Li:2009gj} indicate that
this character is corresponding to an asymptotically stable focus.
And the related  analytical discussion  in \cite{Ao:2010mg} confirmed this conclusion.
However, the researches mentioned above only presented some qualitative results,
where parameters $a_{2}$ and $b$ and initial values are set to certain particular positive values
by hand rather than the values constrained via the observational data, and thus, quantitative analyses are still needed to be done.
 Therefore, in order to  investigate the evolution of the real universe,
it is necessary to place the constraints of parameters obtained via the SNeIa
data on this model.

\begin{figure}
\centering
\epsfig{file=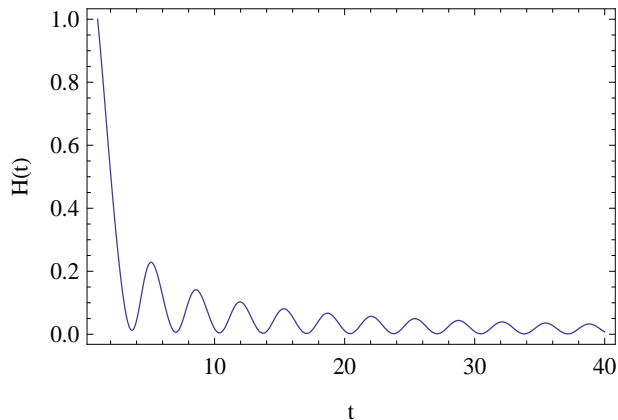,height=2.2in,width=3.2in}
\caption{\label{fig:best-fit}The late-time evolution of torsion cosmology with
  the best-fit values of model parameters and initial condtions for Union 2 SNeIa dataset}
\end{figure}

\begin{figure}[h]
\centering
\includegraphics[width=7cm]{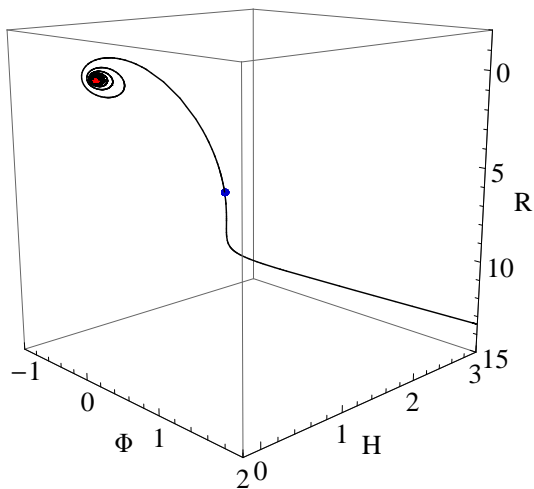}
\caption{\label{fig:evolution}The evolution orbit of ($H,\Phi,R$) with the best-fit values. The blue point indicates the
initial condition (1,~0.584,~5.839), whereas the red point (0,~0,~0) denotes the final state of our universe.}
\end{figure}

\begin{figure}
\centering
\epsfig{file=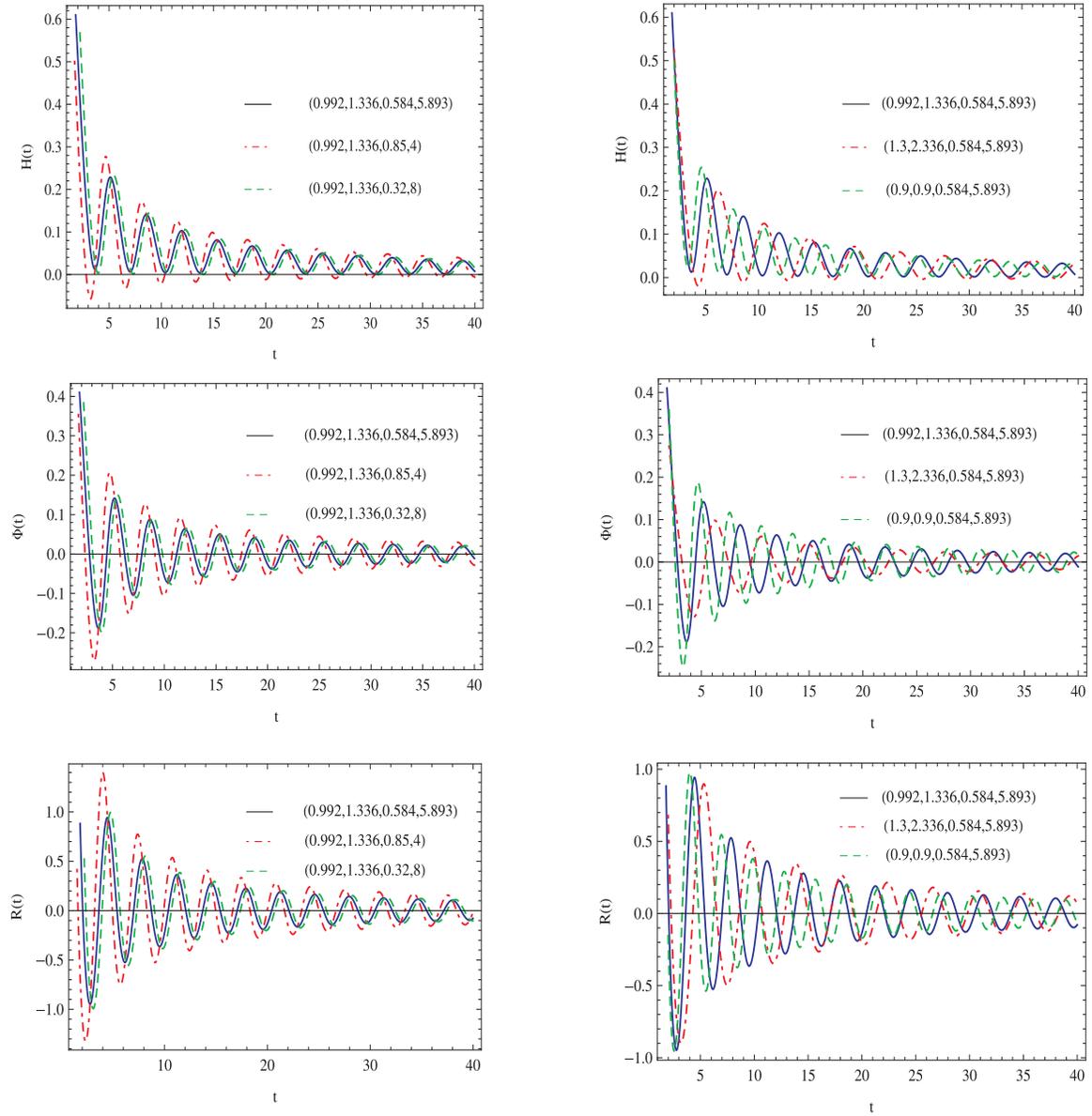,height=6.2in,width=6.0in}
\caption{\label{fig:best-fit-model-initial}The late-time evolution of PGT cosmology with various values of model parameters and
initial conditions lying in the confidence interval of $3 \sigma$. In the left column, the model parameters are fixed at their best-fit values and the initial conditions are varied, whereas in the right column, the initial conditions are fixed at their best-fit values and the model parameters are changed.}
\end{figure}

First we set  model parameters and intial conditions to their best-fit values,
and solve the evolution equations numerically. The solution of Hubble
parameter and the trajectory in $H-\Phi-R$ space are plotted in Fig.~\ref{fig:best-fit} and
Fig.~\ref{fig:evolution}, respectively.
Then we test some other values of model parameters in the confidence interval of
$3\sigma$ with fixed intial conditions which are still at the best-fit value, and plot the numerical
solutions of $H(t)$ in the right column in Fig.~\ref{fig:best-fit-model-initial}.
At last, we fixed the model parameters at their best-fit value, and solve the
evolution equations numerically with various values of initial conditions in the confidence
interval of $3\sigma$. Then we plot the results in the left column in Fig.~\ref{fig:best-fit-model-initial}.
From these numerical results, it is easy to find that ($H$, $\Phi$, $R$) would
tend to ($0,~0,~0$) in the far future, which indicate that  our universe would expand forever
asymptotically to a halt.  This description of the fate of the universe is in
accordance with the earlier analytical work.\cite{Ao:2010mg}

\section{Summary and Conclusion}
\label{sec:summary}
We studied the cosmology based on PGT with even parity scalar mode
dynamical connection,  $0^{+}$. So the Lagrangian we use in this paper takes
the form of SNY model Eq.\eqref{eq:lagrangian-2}. We rewrote the motion
equations Eqs.\eqref{eq:motion-equation-H}-Eqs.\eqref{eq:motion-equation-R} as
a dimensionless equation w.r.t scale factor $a$ rather than cosmological time. And
we obtained the analytical solution of past time evolution of  this new set of
equations, as shown in Eqs.\eqref{eq:ansatz-h}-\eqref{eq:ansatz-r}. Then we
attempted to investigate the goodness of this model, by comparing these
theoretical results with the latest observational data, Union 2 SNeIa dataset.
Finally, we found the best-fit values of model parameters and initial
conditions ($a_{2}=1.336,~b=0.992,$ $ ~\varphi_{0}=0.584$ and $r_{0}=5.839$),
and that the associated minimal $\chi^{2}$ (535.284) is consistent with the
$\Lambda$CDM at the 1$\sigma$ level. Furthermore, from the contours of
the dynamics analysis conducted in some confidence level Fig.\ref{fig:mpcz} and \ref{fig:ivcz}, it
is easy to see that, to some extent, the fine-tuning problem has been alleviated
in SNY model.
Next, we extended our investigation to future evolution. We plotted the whole
evolution orbit of ($H,\Phi,R$), from the past to the future, with the best-fit
values, in Fig. \ref{fig:evolution}, which gives us a raw picture of the whole evolution.
Finally, we tested some other values of parameters and initial conditions and
found that $H,\Phi,R$ all tend to zero in the infinite future, which indicate
our universe will expand forever asymptotically to a halt.
This description of the fate of our universe is
consistent with the analysis conducted above and the earlier works \cite{Li:2009zzc,Li:2009gj, Ao:2010mg}.
Thus, torsion cosmology is a "competitive" model to explain the cosmic
acceleration, which need not introduce some exotic matter composition.

In comparison to other models of accelerating universe, torsion cosmology of PGT is new, which
still has a great number of issues to study. For instance, we could extend the
research on SNY model to BHN model, which generalizes SNY model
to a model with both  $0^{\pm}$ and even the
coupled term of these two modes. Also, we could investigate the effect of
 $0^{-}$ in the very early universe, which might has imprints in  CMBR.
These issues will considered in the upcoming papers.

\begin{acknowledgments}
We wish to acknowledge the support of the SRFDP under Grant No 200931271104 and Shanghai Natural Science
  Foundation, China Grant No. 10ZR1422000.
\end{acknowledgments}

\bibliographystyle{jhep}
\bibliography{pgtc}

\end{document}